\documentclass[aps,prd,draft,showpacs]{revtex4}

\begin{document}

\title{RG Invariance of the Pole Mass in the Minimal Subtraction Scheme}
\author{Chungku Kim}
\date{\today}

\begin{abstract}
We prove the renormalization group(RG) invariance of the pole mass with
respect to the RG functions of the minimal subtraction(MS) scheme and
illustrate this in case of the the neutral scalar field theory both in the
symmetric and in the broken symmetry phase.
\end{abstract}
\maketitle

\affiliation{Department of Physics, Keimyung University, Daegu 704-701,Korea}

%%\pacs{11.15.Bt, 12.38.Bx} after abstract

\affiliation{Department of Physics, College of Natural Science, Keimyung
University, Daegu 705-701, KOREA}

\section{Introduction}

The pole mass defined as the pole of the propagator plays an important role
in the process where the characteristic scale is close to the mass shell\cite
{Narison-1}. The one-loop relation between the scalar quartic coupling
constant and the pole mass of the standard model was obtained long ago\cite
{Sirlin} and the two-loop pole mass of the standard model has been obtain
later\cite{Jegerlehner}. Recently, it was shown that the pole mass is
infrared finite and gauge invariant\cite{Kronfeld}. The RG evolution of the
pole mass in the perturbative calculation was used to determine the bound of
the Higgs mass\cite{Kielanowski}. In this paper, we will prove that the pole
mass when expressed in terms of the Lagrangian parameters, satisfies the RG
equation with respect to the RG functions of the MS scheme both in the
symmetric phase and in broken symmetry phase. In case of the broken symmetry
phase $(m^{2}<0)$, we will follow an approach where we first determine the
VEV as a function of the Lagrangian parameters from the effective
potential(EP) in the symmetric phase ($m^{2}>0)$ renormalized in the MS
scheme. Then by shifting the field to one with vanishing VEV, we obtain the
Lagrangian in broken symmetry phase which contains both the usual $\frac{1}{%
\varepsilon }$ divergent counter-terms and the finite counter-terms coming
from the $O(\hbar )$ terms of the VEV's. In contrast to the usual on shell
renormalization scheme where the VEV is introduced as a free parameter and
is fixed from the no tadpole condition, by treating the $O(\hbar ^{n})$ $%
(n>1)$ quantity which appears in the broken symmetry phase of the Lagrangian
due to the VEV as a finite counter-terms, the tadpole terms vanish
automatically\cite{Taylor} in our approach. In Sec.II, we first prove the RG
invariance of the pole mass starting from the RG equation in the MS scheme
and obtain the pole mass in the symmetric and broken symmetry phase. The
resulting pole mass in the symmetric and broken symmetry phase satisfies the
RG equation with the RG functions of the symmetric phase as shown recently
in case of the EP\cite{Kim}. In Sec.III, we give some discussions and
conclusions.

\section{RG Invariance of the Pole Mass}

In this section, we will prove the RG invariance of the pole mass of the
neutral scalar field theory and demonstrate it explicitly by calculating the
pole mass both in the symmetric and in broken symmetry phase. The
generalization to the other cases will be straightforward. The renormalized
effective action $\Gamma _{R}^{MS}[\phi ]$ of the neutral scalar field
theory in the minimal subtraction scheme(MS) is independent of the
renormalization mass scale $\mu $ and satisfies the RG equation\cite{RG} 
\begin{equation}
(D+\gamma ^{MS}\phi _{z}\frac{\delta }{\delta \phi _{z}})\Gamma
_{R}^{MS}(\mu ,\lambda ,m^{2},\phi )=0,
\end{equation}
where 
\begin{equation}
D\equiv \mu \frac{\partial }{\partial \mu }+\beta _{\lambda }^{MS}\frac{%
\partial }{\partial \lambda }+\beta _{m^{2}}^{MS}\frac{\partial }{\partial
m^{2}},
\end{equation}
and we use the notation that the repeated letters mean the integration over
continuous variables. Here $\beta _{\lambda }^{MS},$ $\beta _{m^{2}}^{MS}$
and $\gamma ^{MS}$ are the RG functions in the MS scheme. In the broken
symmetry phase ($m^{2}<0)$, the scalar field $\phi $ develop a non-vanishing
vacuum expectation value (VEV) $v$ satisfying 
\begin{equation}
\left[ \frac{\delta \Gamma _{R}^{MS}(\mu ,\lambda ,m^{2},\phi )}{\delta \phi 
}\right] _{\phi =v}=0,
\end{equation}
from which one can determine $v$ as a function of $\mu ,\lambda $ and $m^{2}.
$ Then by shifting $\phi \rightarrow \phi +v$, we obtain the effective
action in the broken symmetry phase $\Gamma _{R}^{SB}(\mu ,\lambda
,m^{2},\phi )$ as 
\begin{equation}
\Gamma _{R}^{SB}(\mu ,\lambda ,m^{2},\phi )\equiv \Gamma _{R}^{MS}(\mu
,\lambda ,m^{2},\phi +v(\mu ,\lambda ,m^{2})),
\end{equation}
which satisfy the RG equation with the RG functions as in the symmetric phase%
\cite{Kim} so that 
\begin{equation}
(D+\gamma ^{MS}\phi \frac{\partial }{\partial \phi })\Gamma _{R}^{SB}(\mu
,\lambda ,m^{2},\phi )=0.
\end{equation}
The renormalized N-point one-particle-irreducible (1PI) vertex can be
obtained from the effective action $\Gamma _{R}^{(N)}(\mu ,\lambda
,m^{2},\phi )$ by taking the functional derivative with respect to $\phi $
N-times and putting $\phi =0$ so that 
\begin{equation}
\Gamma _{R}^{(N)}(\mu ,\lambda ,m^{2})=[\frac{\delta ^{N}\Gamma _{R}(\mu
,\lambda ,m^{2},\phi )}{\delta \phi _{x1}\delta \phi _{x2}\cdots \delta \phi
_{xN}}]_{\phi =0}.
\end{equation}
By using Eqs.(1) and (6), one can obtain the RG equation satisfied by $%
\Gamma _{R}^{(N)}(\mu ,\lambda ,m^{2})$ as 
\begin{equation}
(D+N\gamma ^{MS})\Gamma _{R}^{(N)}(\mu ,\lambda ,m^{2})=0.
\end{equation}
Now, the pole mass M is defined as the pole of the Green function $G$ which
is the inverse of $\Gamma _{R}^{(2)}$ so that 
\begin{equation}
\left[ G^{-1}\right] _{p^{2}=-M^{2}}=\left[ p^{2}+\overline{m}^{2}+\Pi
_{R}(p^{2},\mu ,\lambda ,\overline{m}^{2})\right] _{p^{2}=-M^{2}}=-M^{2}+%
\overline{m}^{2}+\Pi _{R}(-M^{2},\mu ,\lambda ,\overline{m}^{2})=0,
\end{equation}
where $\overline{m}^{2}$ is the mass term of the tree level Lagrangian which
becomes $m^{2}$ in the symmetric phase ($m^{2}>0)$ and $-2m^{2}$ in the
broken symmetry phase ($m^{2}<0)$ and $\Pi _{R}(p^{2},\mu ,\lambda ,m^{2})$
is the renormalized self energy obtained from the 1PI two-point Feynman
diagrams. In perturbative calculations, we substitute the expansion of $\Pi
_{R}(p^{2},\mu ,\lambda ,\overline{m}^{2})$ and $M^{2}$ in $\hbar $%
\begin{eqnarray}
\Pi _{R}(p^{2},\mu ,\lambda ,\overline{m}^{2}) &=&\hbar \Pi _{R1}(p^{2},\mu
,\lambda ,\overline{m}^{2})+\hbar ^{2}\Pi _{R2}(p^{2},\mu ,\lambda ,%
\overline{m}^{2})+\cdot \cdot \cdot ,  \nonumber \\
M^{2} &=&M_{0}^{2}+\hbar M_{1}^{2}+\hbar ^{2}M_{2}^{2}+\cdot \cdot \cdot ,
\end{eqnarray}
into Eq.(8) and obtain $M^{2}$ as a function of $\mu ,\lambda $ and $%
\overline{m}^{2}$ as 
\begin{equation}
M^{2}(\mu ,\lambda ,\overline{m}^{2})=\overline{m}^{2}+\hbar \left[ \Pi
_{R1}(p^{2},\mu ,\lambda ,\overline{m}^{2})\right] _{p^{2}=-\overline{m}%
^{2}}+\hbar ^{2}\left[ \Pi _{R2}(p^{2},\mu ,\lambda ,\overline{m}^{2})-\Pi
_{R1}(p^{2},\mu ,\lambda ,\overline{m}^{2})\frac{\partial \Pi
_{R1}(p^{2},\mu ,\lambda ,\overline{m}^{2})}{\partial p^{2}}\right] _{p^{2}=-%
\overline{m}^{2}}+O(\hbar ^{3}).
\end{equation}
Now, the RG equation for $G^{-1}$ can be obtained by putting N=2 in Eq.(7)
as 
\begin{equation}
0=(D+2\gamma ^{MS})(p^{2}+\overline{m}^{2}+\Pi _{R}(p^{2},\mu ,\lambda ,%
\overline{m}^{2}))=D(\overline{m}^{2}+\Pi _{R}(p^{2},\mu ,\lambda ,\overline{%
m}^{2}))+2\gamma ^{MS}(p^{2}+\overline{m}^{2}+\Pi _{R}(p^{2},\mu ,\lambda ,%
\overline{m}^{2})).
\end{equation}
By applying the operation $D$ to the second equation of Eq.(8), we obtain 
\begin{equation}
0=D(-M^{2}+\overline{m}^{2}+\Pi _{R}(-M^{2},\mu ,\lambda ,\overline{m}%
^{2}))=(DM^{2})(-1+\frac{\partial \Pi _{R}(-M^{2},\mu ,\lambda ,\overline{m}%
^{2})}{\partial M^{2}})+[D(p^{2}+\overline{m}^{2}+\Pi _{R}(p^{2},\mu
,\lambda ,\overline{m}^{2})]_{p^{2}=-M^{2}}.
\end{equation}
By substituting Eq.(11) into (12), we obtain 
\begin{equation}
0=(DM^{2})(-1+\frac{\partial \Pi _{R}(-M^{2},\mu ,\lambda ,\overline{m}^{2})%
}{\partial M^{2}})-[2\gamma ^{MS}(p^{2}+\overline{m}^{2}+\Pi _{R}(p^{2},\mu
,\lambda ,\overline{m}^{2})]_{p^{2}=-M^{2}}.
\end{equation}
The last term of the of the above equation vanishes due to the definition of
the pole mass (Eq.(8)) and as a result, we obtain 
\begin{equation}
DM^{2}=0,
\end{equation}
which means that the pole mass is RG invariant.

Now let us consider the case of the neutral scalar field theory with the
Euclidean classical action

\begin{equation}
S[\phi ]=\int d^{4}x(\frac{1}{2}Z_{\phi }\phi (-\partial ^{2}+m_{B}^{2})\phi
+\frac{1}{24}\lambda _{B}Z_{\phi }^{2}\phi ^{4}).
\end{equation}
Up to two-loop, the renormalized self energy $\Pi _{R}(p^{2},\mu ,\lambda
,m^{2})$ of the neutral scalar field theory is given by 
\begin{equation}
\Pi _{R}(p^{2},\mu ,\lambda ,m^{2})=\begin{picture}(220,20)
\put(3,2){\line(1,0){16}} \put(8,0) {x} \put(30,0){+ $\frac{1}{2}$}
\put(50,0){ \line(1,0){16}} \put(60,8){\circle{16}} \put(80,0) {+
$\frac{1}{2}$} \put(100,0){ \line(1,0){16}} \put(110,8){\circle{16}}
\put(104,-1) {\makebox(4,4)} \put(130,0){+ $\frac{1}{2}$} \put(150,0){
\line(1,0){16}} \put(160,8){\circle{16}} \put(157,14) {x} \put(180,0){+
$\frac{1}{8}$} \put(200,5){\line(1,0){28}} \put(212,5){\circle{16}}
\put(235,0){- $\frac{1}{12}$} \put(265,5){\circle{14}}
\put(265,19){\circle{14}} \put(255,-2){\line(1,0){20}} 
\put(285,0){,}\end{picture}
\end{equation}
where the cross in the line means the counter-term for the propagator $%
(Z_{\phi }-1)p^{2}+Z_{\phi }m_{B}^{2}-m^{2}$ and the box means the vertex
counter-term $\lambda _{B}Z_{\phi }^{2}-\lambda $. By using the well known
MS counter-terms given by 
\begin{eqnarray}
\lambda _{B} &=&\mu ^{2\varepsilon }\{\lambda +\frac{\hbar }{(4\pi )^{2}}%
\frac{3\lambda ^{2}}{2\varepsilon }+\frac{\hbar ^{2}}{(4\pi )^{4}}(-\frac{%
\lambda ^{2}}{2\varepsilon ^{2}}+\frac{\lambda ^{2}}{2\varepsilon ^{2}}%
)\cdot \cdot \cdot \},  \nonumber \\
m_{B}^{2} &=&m^{2}\{1+\frac{\hbar }{(4\pi )^{2}}\frac{\lambda }{2\varepsilon 
}+\frac{\hbar ^{2}}{(4\pi )^{4}}(-\frac{5\lambda ^{2}}{24\varepsilon }+\frac{%
\lambda ^{2}}{2\varepsilon ^{2}})+\cdot \cdot \cdot \},  \nonumber \\
Z_{\phi } &=&\mu ^{-\varepsilon }\{1-\frac{\hbar ^{2}}{(4\pi )^{4}}\frac{%
\lambda ^{2}}{24\varepsilon ^{2}}\cdot \cdot \cdot \},
\end{eqnarray}
we can obtain 
\begin{eqnarray}
\Pi _{R1}(p^{2},\mu ,\lambda ,m^{2}) &=&\frac{\lambda m^{2}}{32\pi ^{2}}%
\{\ln (\frac{m^{2}}{\mu ^{2}})-1\},  \nonumber \\
\Pi _{R2}(p^{2},\mu ,\lambda ,m^{2}) &=&\frac{\lambda ^{2}m^{2}}{(16\pi
^{2})^{2}}\{\frac{1}{2}\ln ^{2}(\frac{m^{2}}{\mu ^{2}})-\frac{5}{4}\ln (%
\frac{m^{2}}{\mu ^{2}})+\frac{1}{12}p^{2}\ln ^{2}(\frac{m^{2}}{\mu ^{2}})-%
\frac{1}{6}J_{3}(\frac{m^{2}}{p^{2}})\},
\end{eqnarray}
where $J_{3}(\frac{m^{2}}{p^{2}})$ can be found in \cite{Fleischer}. By
using(10) one can obtain the expansion of the pole mass $M^{2}$ up to the
order $\hbar ^{2}$ as 
\begin{equation}
M^{2}=m^{2}+\frac{\hbar }{32\pi ^{2}}\lambda m^{2}\{(\ln (\frac{m^{2}}{\mu
^{2}})-1\}+\frac{\hbar ^{2}\lambda ^{2}}{(16\pi ^{2})^{2}}\lambda ^{2}m^{2}\{%
\frac{1}{2}\ln ^{2}(\frac{m^{2}}{\mu ^{2}})-\frac{7}{6}\ln (\frac{m^{2}}{\mu
^{2}})-\frac{1}{6}J_{3}(1)\}\cdot \cdot \cdot .
\end{equation}
By using the RG functions in MS scheme\cite{RG} 
\begin{eqnarray}
\beta _{\lambda }^{MS} &=&\mu \frac{d\lambda }{d\mu }=3\frac{\hbar }{(4\pi
)^{2}}\lambda ^{2}-\frac{17}{3}\frac{\hbar ^{2}}{(4\pi )^{4}}\lambda ^{3}+(%
\frac{145}{8}+12\text{ }\varsigma (3))\frac{\hbar ^{3}}{(4\pi )^{6}}\lambda
^{4}+\cdot \cdot \cdot , \\
\beta _{m^{2}}^{MS} &=&\frac{\mu }{m^{2}}\frac{dm^{2}}{d\mu }=\frac{\hbar }{%
(4\pi )^{2}}\lambda -\frac{5}{6}\frac{\hbar ^{2}}{(4\pi )^{4}}\lambda ^{2}+%
\frac{7}{2}\frac{\hbar ^{3}}{(4\pi )^{6}}\lambda ^{3}+\cdot \cdot \cdot , \\
\gamma ^{MS} &=&\frac{\mu }{\phi }\frac{d\phi }{d\mu }=-\frac{1}{12}\frac{%
\hbar ^{2}}{(4\pi )^{4}}\lambda ^{2}+\frac{1}{16}\frac{\hbar ^{3}}{(4\pi
)^{6}}\lambda ^{3}+\cdot \cdot \cdot ,
\end{eqnarray}
we can confirm the RG invariance of the pole mass $M^{2}$ given in Eq.(19).

Now, let us consider the case of broken symmetry phase ( $m^{2}<0)$. The
Euclidean action for the broken symmetry phase can be obtained from that of
the symmetric phase by shifting the field $\phi \longrightarrow \phi +v.$ In
case of the neutral scalar field theory, $\ $the perturbative expansion of $v
$ can be obtained as\cite{Kim} 
\begin{equation}
v(\mu ,\lambda ,m^{2})=v_{0}(\lambda ,m^{2})+\hbar v_{1}(\mu ,\lambda
,m^{2})+\cdot \cdot \cdot ,
\end{equation}
where 
\begin{equation}
v_{0}^{2}=-m^{2}\frac{6}{\lambda }\text{ and }v_{0}v_{1}=\frac{3}{16\pi ^{2}}%
\lambda m^{2}\{\ln (\frac{-2m^{2}}{\mu ^{2}})-1\}.
\end{equation}
As a result, the $O(\hbar )$ terms of the classical action contains both the
usual $\frac{1}{\varepsilon }$ divergent counter-terms and the finite
counter-terms coming from the $O(\hbar )$ terms of the VEV's so that at
one-loop we obtain

\begin{eqnarray}
S[\phi ] &=&\int d^{4}x[\frac{1}{2}\phi (-\partial ^{2}+m^{2}+\frac{1}{2}%
\lambda v_{0}^{2})\phi +\frac{1}{6}\lambda v_{0}\phi ^{3}+\frac{1}{24}%
\lambda \phi ^{4}+(Z_{\phi }m_{B}^{2}v+\frac{1}{6}\lambda _{B}Z_{\phi
}^{2}v^{3})\phi   \nonumber \\
&&+((Z_{\phi }-1)p^{2}+Z_{\phi }m_{B}^{2}-m^{2}+\frac{1}{2}(\lambda
_{B}Z_{\phi }^{2}v^{2}-\lambda v_{0}^{2}))\phi ^{2}+\frac{1}{6}(\lambda
_{B}Z_{\phi }^{2}v-\frac{\lambda }{6}v_{0})\phi ^{3}+(\lambda _{B}Z_{\phi
}^{2}-\lambda )\phi ^{4}]  \nonumber \\
&=&\int d^{4}x[\frac{1}{2}\phi (-\partial ^{2}-2m^{2})\phi +\frac{1}{6}%
\lambda v_{0}\phi ^{3}+\frac{1}{24}\lambda \phi ^{4}+\hbar \{(-\frac{1}{%
(4\pi )^{2}}\frac{\lambda }{\varepsilon }v_{0}-2v_{1})m^{2}\phi   \nonumber
\\
&&+(-\frac{1}{(4\pi )^{2}}\frac{2\lambda }{\varepsilon }m^{2}+\frac{1}{2}%
\lambda v_{0}v_{1})\phi ^{2}+(\frac{1}{(4\pi )^{2}}\frac{\lambda ^{2}}{%
4\varepsilon }v_{0}+\frac{1}{6}\lambda v_{1})\phi ^{3}+\frac{1}{(4\pi )^{2}}%
\frac{\lambda ^{2}}{16\varepsilon }\phi ^{4}\}+O(\hbar ^{2}),
\end{eqnarray}
where we have used Eq.(17) to obtain the one-loop counter-terms given in the
last line. The one-loop two-point function is given by 
\begin{equation}
\Pi _{R1}(p^{2},\mu ,\lambda ,m^{2})=\begin{picture}(270,20)
\put(5,2){\line(1,0){16}} \put(10,0) {x} \put(35,0) {+ $\frac{1}{2}$}
\put(55,0){ \line(1,0){16}} \put(65,8){\circle{16}} \put(85,0) {-
$\frac{1}{2}$} \put(110,5){\line(1,0){8}} \put(126,5){\circle{16}}
\put(135,5){\line(1,0){8}} \put(150,0){+$\frac{1}{2}$}
\put(170,0){\line(1,0){16}} \put(178,0){\line(0,1){5}}
\put(178,13){\circle{16}} \put(205,0){+}
\put(215,0){\line(1,0){16}}\put(223,0){\line(0,1){10}}
\put(223,10){\circle*{4}}\end{picture},
\end{equation}
where the four-point vertex has the vertex factor $\lambda $, the
three-point vertex has the vertex factor $\lambda v_{0},$ the filled circle
means the tadpole counter-term $Z_{\phi }m_{B}^{2}v+\frac{1}{6}\lambda
_{B}Z_{\phi }^{2}v^{3}$ and the cross in the line means the counter-term for
the propagator $(Z_{\phi }-1)p^{2}+Z_{\phi }m_{B}^{2}-m^{2}+\frac{1}{2}%
(\lambda _{B}Z_{\phi }^{2}v^{2}-\lambda v_{0}^{2})$. As noted in
introduction, the last two terms of the above equation which are the
one-loop tadpole terms cancel out exactly and the sum of the first three
terms of the above equation give finite result as 
\begin{equation}
\Pi _{R1}(p^{2},\mu ,\lambda ,m^{2})=\frac{2\lambda m^{2}}{16\pi ^{2}}(\ln (%
\frac{-2m^{2}}{\mu ^{2}})+1)-\frac{3\lambda m^{2}}{16\pi ^{2}}%
\int_{0}^{1}d\alpha \text{ }\ln \{\frac{-2m^{2}+\alpha (1-\alpha )p^{2}}{\mu
^{2}}\}.
\end{equation}
Note that the $\frac{{1}}{{\varepsilon }}$ divergences of the two-point
Green function in the broken symmetry phase cancel out by the counter-terms
of the symmetric phase given in Eq.(17)\cite{Symannzik} and that the
two-point Green function in the broken symmetry phase satisfies the RG
equation with the RG functions of the symmetric phase given in Eq.(14). The
pole mass in the broken symmetry phase can be obtained from Eq.(10) as 
\begin{equation}
M^{2}=-2m^{2}+\frac{\hbar }{16\pi ^{2}}\lambda m^{2}\{{-\text{ }\ln (\frac{%
-2m^{2}}{\mu ^{2}})+8-\sqrt{3}\pi \}+O(\hbar ^{2}),}
\end{equation}
which satisfy the RG equation given in Eq.(14) with the RG functions of the
symmetric phase.

From the RG invariance of the pole mass, we can use the method of
characteristics\cite{Ford} to obtain the RG improvement of the pole mass in
the perturbative calculations as 
\begin{equation}
M^{2}(\mu ,\lambda ,m^{2})=M^{2}(\mu e^{2t},\lambda (t),m^{2}(t)),
\end{equation}
where $\lambda (t)$ and $m^{2}(t)$ satisfies 
\begin{eqnarray}
\frac{d\lambda (t)}{dt} &=&\beta _{\lambda }^{MS}(\lambda (t)),  \nonumber \\
\frac{dm^{2}(t)}{dt} &=&\beta _{m^{2}}^{MS}(\lambda (t)),
\end{eqnarray}
with the initial conditions $\lambda (0)=\lambda $ and $m^{2}(0)=m$.

\section{ Discussions and Conclusions}

In this paper, we have proved the RG invariance of the pole mass and have
verified it explicitly from perturbative calculations in case of the scalar
field theory both in the symmetric and in broken symmetry phase. In case of
the broken symmetry phase, we have used an approach where the VEV was
obtained from the EP in the symmetric phase as a function of the Lagrangian
parameters and then the Lagrangian in broken symmetry phase was obtained by
shifting the scalar field in contrast to the usual on shell renormalization
scheme where the VEV introduced as a free parameter in the Lagrangian of the
broken symmetry phase, is fixed by the no tadpole condition. As a result, we
have shown that the resulting pole mass both in the symmetric and in broken
symmetry phase satisfies the RG equation with RG function of the symmetric
phase. The RG invariance of the pole mass in the broken symmetry phase by
using the RG functions of the symmetric phase can be used as a check to the
higher order calculation of the pole mass as well as RG improvement of the
pole mass.

\begin{acknowledgements}
This research was supported in part by the Institute of Natural Science.
\end{acknowledgements}

\bibliographystyle{plain}
\bibliography{2PI}

\begin{thebibliography}{99}
\bibitem{Narison-1}  Narison, see M. Sher, Phys. Rep. 179, 273 (1989).

\bibitem{Sirlin}  A. Sirlin and R. Zucchini, Nucl. Phys. B266 (1986) 389.

\bibitem{Jegerlehner}  F. Jegerlehner, M. Yu. Kalmykov and O. L. Veretin,
Nucl. Phys. B641 (2002) 285; B658 (2003) 49.

\bibitem{Kronfeld}  A. S. Kronfeld, Phys. Rev. D58 (1998) 051501.

\bibitem{Kielanowski}  P. Kielanowski and S. R. Juarez W, Phys. Rev. D72
(2005) 096003.

\bibitem{Taylor}  J. C. Taylor, \emph{Gauge Theories of Weak Interactions}
(Cambridge Press, Cambridge, 1976).

\bibitem{Kim}  C. Kim, Joul. of the Kor. Phys. Soc. 59 (2011) 2993.

\bibitem{RG}  H. Kleinert, J. Neu, V. Schulte-Frohlinde, K. G. Chetyrkin, S.
A. Larin, Phys. Lett. B\textbf{272}, 39(1991); ibid. 319, 545(1993).

\bibitem{Fleischer}  J. Fleischer, F. Jegerlehner,O. V. Tarasov and O. L.
Veretin , Nucl. Phys.B539 (1999) 671

\bibitem{Symannzik}  K. Symanzik, Proc. of ''Fundamental Interactionsat High
Energies'' ed. A. Perlmutter (Gordon and Breach, New York, 1970).

\bibitem{Ford}  C. Ford, D. R.T. Jones, P. W. Stephenson and M. B. Einhorn,
Nucl. Phys. B395 (1993) 17.
\end{thebibliography}

\end{document}